\documentclass[%
 reprint, superscriptaddress,
 amsmath,amssymb,
 aps,
]{revtex4-2}

\usepackage{graphicx}
\usepackage[dvipsnames]{xcolor}
\usepackage{dcolumn}
\usepackage{bm}
\usepackage{bbold}
\usepackage[normalem]{ulem}
\usepackage{dsfont}
\usepackage{subfigure}

\usepackage{physics}
\usepackage{braket, amsmath, xfrac, hyperref, siunitx}
\hypersetup{
    colorlinks,
    citecolor=blue,
    filecolor=blue,
    linkcolor=blue,
    urlcolor=blue
}
\usepackage{natbib}
\usepackage{cancel}
\newcommand{\ba}{\begin{eqnarray}}
\newcommand{\ea}{\end{eqnarray}}
\newcommand{\ban}{\begin{eqnarray*}}
\newcommand{\ean}{\end{eqnarray*}}

\begin{document}

\title{Possibility of detecting gravity of an object frozen in a spatial superposition by the Zeno effect}

\author{Peter Sidajaya}
\affiliation{center for Quantum Technologies, National University of Singapore, 3 Science Drive 2, Singapore 117543}

\author{Wan Cong}
\affiliation{University of Vienna, Faculty of Physics, Boltzmanngasse 5, A 1090 Vienna, Austria} 

\author{Valerio Scarani}
\affiliation{center for Quantum Technologies, National University of Singapore, 3 Science Drive 2, Singapore 117543}
\affiliation{Department of Physics, National University of Singapore, 2 Science Drive 3, Singapore 117542}

\date{\today}

\begin{abstract}


While quantum probes surely feel gravity, no \textit{source} of gravity has been prepared in a delocalized quantum state yet. Two basic questions need to be addressed: how to delocalize a mass sufficiently large to generate \textit{detectable} gravity and, once that state has been prepared, how to \textit{fight localization} by decoherence. We propose to fight decoherence by freezing the source in the desired state through the \textit{Zeno effect}. Successful implementation can be verified by \textit{scattering} a probe in the effective potential generated by the source. Besides putting forward the idea, we provide an estimation of the values of the parameters required for the proposal to be feasible. Overall, the proposal seems as challenging as other existing ones, although the specific challenges are different (e.g.~no entanglement needs to be preserved or detected, but the Zeno freezing must be implemented).

\end{abstract}

\maketitle

\section{Introduction}

\begin{figure}[t]
    \centering
    \includegraphics[width=\linewidth]{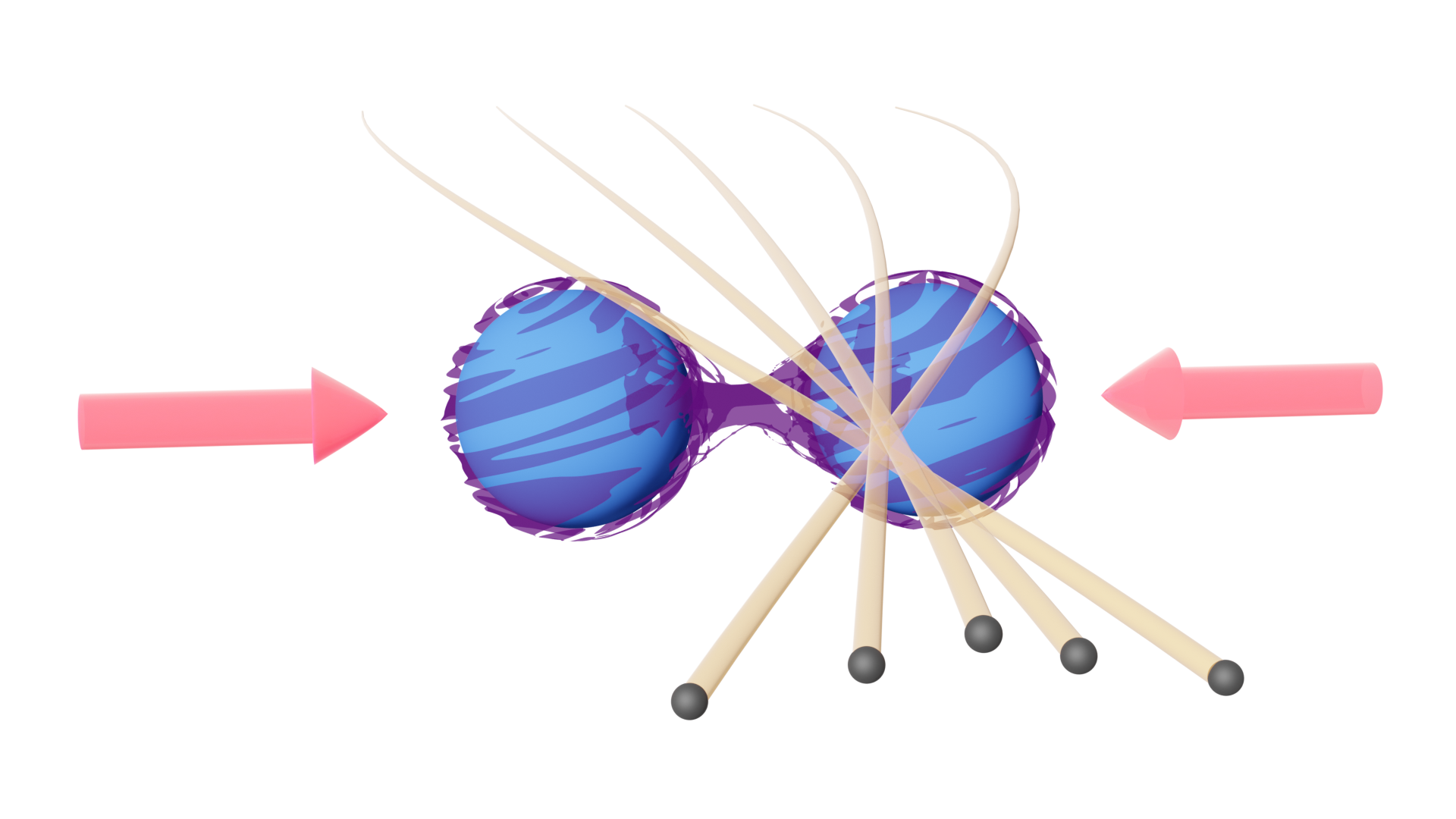}
    \caption{Sketch of the proposal. A source mass is held in superposition (the blue spheres represent the wave function of a single object) by Zeno measurements (red arrow) and smaller test masses (the gray spheres) being scattered by the source masses. The scattering feature of the test masses would help us infer that the gravitational source is in a superposed state.}
    \label{fig:impressive_graphic}
\end{figure}

The interplay between quantum physics and gravity is a source of conceptual challenges for physicists, hopefully to be submitted to observational or experimental tests. On one end of the spectrum of questions, the reconciliation of the two theories in the high energy regime has been elusive (see e.g. \cite{Carlip:2001wq,Rovelli:2000aw}). On the other end, it has never been in doubt that the dynamics of quantum systems is affected by gravity. This has been reported in several types of experiments over the last few decades. For example, the gravitational phase shift of neutrons was measured in \cite{Colella:1975dq}, quantum bound states under Earth's gravitational potential were observed in \cite{Nesvizhevsky:2002ef,Nesvizhevsky:2003ww}, and various atom interferometry experiments under the influence of Earth's gravitational field have been conducted \cite{asenbaum,fray04,Rosi17,Tarallo14}. Coming from this end, the next challenge consists of engineering a \textit{source of gravity} in a quantum state, typically one where the center of mass is strongly delocalized \cite{Carney_review, Huggett:2022uui}.

Detection of the gravitational field created by a source in a quantum state will require enormous technological progress, both in gravity detection and in the quantum control of large systems. Current state-of-the-art techniques allow for the detection of the gravity of masses down to $10^{-4}\,\si{\kg}$ \cite{Westphal_Cavendish}. On the other hand, levitated systems seem to be the system of choice to prepare spatial superpositions \cite{levitoreview}: in that setting, the largest objects that have been cooled to the quantum ground state have a mass of the order of $10^{-17}\,\si{\kg}$ \cite{Deli2020}. Somewhere between these values, we can hope that gravity detection and quantum control will meet. Meanwhile, this no-man's-land is slowly being populated by theoretical proposals on how such states could be created and detected \cite{Derakhshani_2016,Carlesso:2017vrw,Lindner05}. Literature surrounding similar table-top experiments has recently seen a small explosion after the proposals in \cite{Bose17,Marletto:2017kzi}, which considered creating two delocalized quantum systems, and observing entanglement due to gravitational interaction \cite{Marshman:2019sne,Marletto:2020cdx,Galley:2020qsf,Belenchia:2018szb,danielson2022gravitationally}. While there are limits to what such proposals can achieve (for instance, some quantum gravity theories can be differentiated only at Planck-scale energies \cite{Anastopoulos_2020}), they have succeeded in drawing the communities' attention to technological maturity in preparing such systems.

Since \cite{Bose17,Marletto:2017kzi}, numerous table top experiments for observing non classical sources of gravity involving, for example, quantum oscillators and Bose-Einstein condensates have been proposed \cite{Carlesso_2019,Matsumura20,Krisnanda20,Carney21,streltsov2022significance,Howl21,haine2021searching,Weiss21,cosco2021enhanced}. In all the existing proposals, the main obstacle is decoherence \cite{aspelmeyer2022avoid}. Indeed, the more delocalized a system is, the more likely it is that the environment correlates with the position and destroys the coherent superposition. Here, we propose to \textit{fight decoherence by freezing the state of the source object through the Zeno effect}. This freezing prevents that object from getting entangled with anything, including the environment and other controlled systems. Thus, we are moving away from the idea of probing quantumness through entanglement that was central to previous proposals. Instead, our goal is to show that the effective gravitational potential generated by a source frozen in a delocalized state can in principle be detected by scattering a probe. We provide a feasibility estimate for the parameters required to demonstrate such an effect.


\section{The proposal at a glance}
\label{sec:glance}

\begin{figure*}
    \centering
    \includegraphics[width=0.9\linewidth]{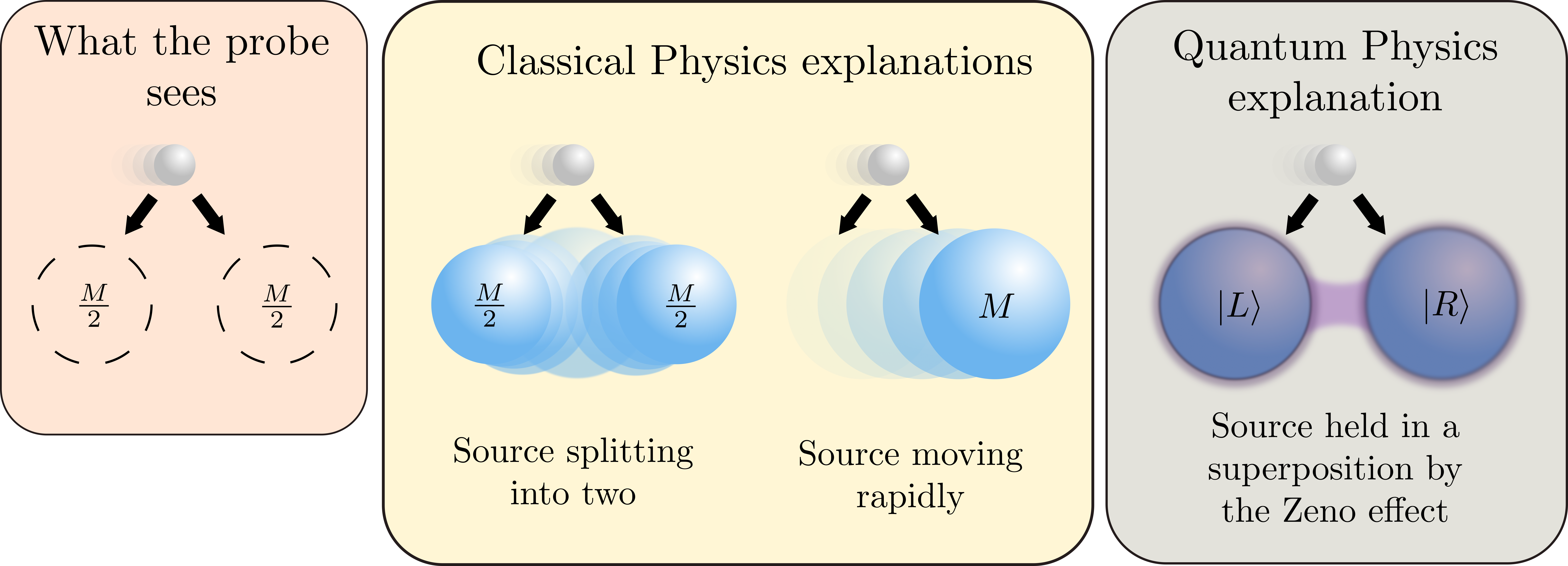}
\caption{A single object (sphere) of mass $M$ is inserted into the experimental apparatus; but the scattering of the probe reveals a delocalized distribution of mass, e.g.~two separate sources of mass $M/2$ (\textbf{left}). If the experiment has been designed to Zeno-freeze the source into a quantum superposition of the center of mass, the natural explanation of the observation is that the procedure has been successful (\textbf{right}). The two alternative classical descriptions (\textbf{middle}) are highly contrived: (i) the source splits into two smaller masses during the experiment, or (ii) the source is rapidly moving between the two positions in such a fine-tuned way as to generate the same effective potential.}
    \label{fig:argument}
\end{figure*}

Let us first state the overall idea of this proposal. The \textit{source of gravity} is an object of total mass $M$, distributed according to the density $\rho(\mathbf{X}+\mathbf{r})$ around its center of mass $\mathbf{X}$. Our goal is to certify, using gravitational interaction, that the center of mass is delocalized according to the wave-function $\phi(\mathbf{X})$. For this, we consider using a \textit{probe}, which is a material point of mass $m$ and dynamical variables $(\mathbf{x},\mathbf{p})$. The protocol consists of scattering the probe in the gravitational potential of the source, while it is kept frozen in the state $\phi(\mathbf{X})$ by the Zeno effect. An illustration of the proposal can be seen in Fig. \ref{fig:impressive_graphic}.

Under these conditions, we shall show in Sec. \ref{sec:zeno} that the probe experiences unitary scattering by the potential
\ba
V_\phi(\mathbf{x})&=&\int d^3\mathbf{X}\,|\phi(\mathbf{X})|^2\,V(\mathbf{x},\mathbf{X})\,,\label{Veff}
\ea
where 
\ba
V(\mathbf{x},\mathbf{X})&=&-GmM\,\int d^3\mathbf{r}\, \frac{\rho(\mathbf{X}+\mathbf{r})}{|\mathbf{x}-(\mathbf{X}+\mathbf{r})|}\,.\label{Vint}
\ea
Thus, $V_\phi(\mathbf{x})$ is identical to the gravitational potential generated by a mass distribution $M|\phi(\mathbf{X})|^2 \rho(\mathbf{X}+\mathbf{r})$.

This potential appears here as the \textit{exact} prediction for Zeno freezing and not as the result of approximate models like ``semiclassical gravity''. Since the very same potential can be realized in those theories, our proposal cannot exclude them, in contrast to the proposals which rely on entanglement \cite{Carney_review, Huggett:2022uui}.

Even if semi-classical theories cannot be excluded, we argue that \textit{a fully classical description of the experiment is impossible, and therefore, one can conclude that the observation is due to quantum delocalization}. The argument (presented pictorially in Fig.~\ref{fig:argument}) is as follows. There are two possible classical descriptions of the observation. The first description is the obvious one, using a \textit{static} mass distribution. Surely, one can create that distribution, but this is not how the full tabletop controlled experiment will work. There, one inserts the source object with a given shape (say, a sphere) into the experiment chamber at the start, and at the end extracts the object with the same shape. In this context, it is incredible to believe that the matter morphed into a different shape (the shape prescribed by $|\phi(\mathbf{X})|^2$) during the run of the experiment before returning to the original one. The second description is based on \textit{dynamical} delocalization: the object retains its shape, but its center of mass undergoes a motion $\mathbf{X}(t)$ that is so fast and fine tuned that the probe perceives it as the coarse-grained static distribution $|\phi(\mathbf{X})|^2$. Such an explanation is conspiratorial, and in specific examples, it could be conclusively ruled out by explicit classical simulations of trajectories in the potential. In summary, both the static and the dynamical classical descriptions are at the boundary of the unbelievable, and one is left with the explanation based on the quantum delocalization of the source.

The remainder of this paper is structured as follows. In Sec. \ref{sec:zeno}, we discuss the Zeno effect and derive some conditions, under which \eqref{Veff} applies. In Sec. \ref{sec:proposal}, we discuss the proposed experiment and how it would be able to tell the presence of a delocalized source. In Sec. \ref{sec:feasibility}, we put numbers into these conditions and compare them also with expected sources of decoherence and other effects that may interfere with our intended detection.

\section{The Zeno Effect on the gravity source}
\label{sec:zeno}

Since it was proposed by Misra and Sudarshan \cite{Misra:1976by}, the quantum Zeno effect (QZE) has been the object of extensive theoretical work \cite{Peres:1980ux,PRESILLA96,Onofrio:1993uj,Facchi:2000bs,Facchi01,Facchi02,Koshino:2004rw} and of a variety of experimental demonstrations \cite{Itano1990,Itano_2009,Siddiqi16,siddiqi18,Childress06,Wolters13,gagen93,altenmuller94,Fischer01,Patil15,Elliot16,mekhov2012quantum, elliott2015multipartite}. In this section, we first rederive how the QZE on one system affects the dynamics of another system interacting with it. Then we sketch a possible implementation of the QZE suited to our proposal.

\subsection{Effect of freezing the source on the dynamics of the probe}
\label{subsec:zeno}

We consider a system composed of two subsystems, denoted $P$ (probe) and $S$ (source), the latter being the one whose state we want to freeze to $\ket{\phi}$. The system evolves accordingly to the Hamiltonian $H=H_P+H_S+H_{\textrm{int}}$ for a short time $\tau$; then, an instantaneous two-outcome projective measurement is performed on $S$, described by $P_\phi=\ket{\phi}\bra{\phi}$ and $P_{\perp}=\mathbb{1}-P_\phi$. In our proposal, it is natural to assume that the source is trapped in a potential, such that $H_S\ket{\phi}=E\ket{\phi}$. Thus, leaving other sources of decoherence aside for the moment, the Zeno freezing is meant to counter only the evolution induced by $H_{\textit{int}}$.

The stroboscopic evolution on the total system is given by
\ban
\rho((n+1)\tau)&=&\sum_{j=\phi,\perp}\left[\mathbb{1}\otimes P_j\right]\,\mathcal{U}_H[\rho(n\tau)]\,\left[\mathbb{1}\otimes P_j\right],
\ean 
where we denoted
\ban
\mathcal{U}_H[\rho]&=&e^{-iH\tau/\hbar}\rho e^{iH\tau/\hbar}\\
&=&\rho-i\frac{\tau}{\hbar}[H,\rho]-\frac{\tau^2}{2\hbar^2}\left[H,[H,\rho]\right]+O(\tau^3).
\ean
Assume now that system $S$ was prepared in state $\ket{\phi}$, and that Zeno freezing was successful for the first $n$ steps, i.e., $\rho(n\tau)=\alpha(n\tau)\otimes P_\phi$. Then at order $\tau^2$ it holds that
\ba
\rho((n+1)\tau)&=&\mathcal{U}_{H_\phi}[\alpha(n\tau)]\otimes P_\phi\nonumber\\
&&-\frac{\tau^2}{2\hbar^2}\left\{\Delta H^2_{\,\phi},\alpha(n\tau)\right\}\otimes P_\phi\nonumber\\
&&+\frac{\tau^2}{\hbar^2}\,\left[\mathbb{1}\otimes P_\perp\right]H\rho(n\tau)H\left[\mathbb{1}\otimes P_\perp\right]\,, \label{strobo}
\ea
where
\ba
H_\phi&=&\textrm{Tr}_S\left(\left[\mathbb{1}\otimes P_\phi\right]\,H\right),\\
\Delta H^2_{\,\phi}&=& \textrm{Tr}_S\left(\left[\mathbb{1}\otimes P_\phi\right]\,H^2\right)-H_\phi^{\,2}
\ea are operators that act on the probe. Thus, in the infinitesimal limit $\tau\rightarrow 0$, the stroboscopic evolution of the probe-system state at time $t=n\tau$ becomes:
\begin{align*}
\frac{d\rho(t)}{dt} 
= -\frac{i}{\hbar}[H_\phi, \alpha(t)] \otimes P_\phi.
\end{align*}
In other words, the evolutions of $S$ and $P$ are given by
$$
\Tr_P\left[\frac{d\rho(t)}{dt}\right]=0, \qquad \Tr_S\left[\frac{d\rho(t)}{dt}\right]=-\frac{i}{\hbar}[H_\phi, \alpha(t)];
$$
the state of $S$ is exactly frozen in $\ket{\phi}$, while $P$ undergoes a unitary evolution  under $H_\phi$, validating Eq.~\eqref{Veff}. That being said, the infinitesimal limit is mathematical, and we want better control of what it means for the time $\tau$ to be ``short.'' On the one hand, every step must be short compared to the timescale of the interaction. We can use the estimate 
\ba\label{eq:zeno1}
\tau&\ll &\hbar\,\left|-\frac{GmM}{b_0}\right|^{-1},
\ea
where $b_0$ is the impact parameter of the scattering trajectory. The right-hand side comes from an estimate of the magnitude of $\hbar ||H_\phi^2||^{-1/2} \sim \hbar ||\Delta H_\phi^2||^{-1/2}$. This value is also an estimate of what is commonly called the Zeno time $\tau_Z = \hbar ||\Delta H_\phi^2||^{-1/2} $ in the literature. The bound $\tau \ll \tau_Z$, which we use for our preliminary estimates, must certainly be enforced, but may be loose: it will have to be scrutinized for each specific realization \cite{Facchi01}.

Finally, we have to ensure the survival probability of the system for the duration of the experiment. For the superposition to survive for $t_{total}$, during which $N$ measurements are made, the survival probability 
\begin{equation*}
    p^{(N)}(t_{total}) \approx \left[ 1 - \left( \frac{\tau}{\tau_Z} \right)^2\right]^N \approx 1 - N \left( \frac{\tau}{\tau_Z} \right)^2
\end{equation*}
must be reasonably large $p^{(N)} \sim 1$. This will yield 
\begin{equation}\label{eq:zeno2}
    t_{total} \ll \frac{\tau_Z^2}{\tau}\,.
\end{equation}
In our context, this bounds the time allotted to the probe to feel the gravitational effect of the source.

\subsection{Case study: The source state as the ground state of suitable potentials}

\begin{figure}
    \centering
    \includegraphics[width=\linewidth]{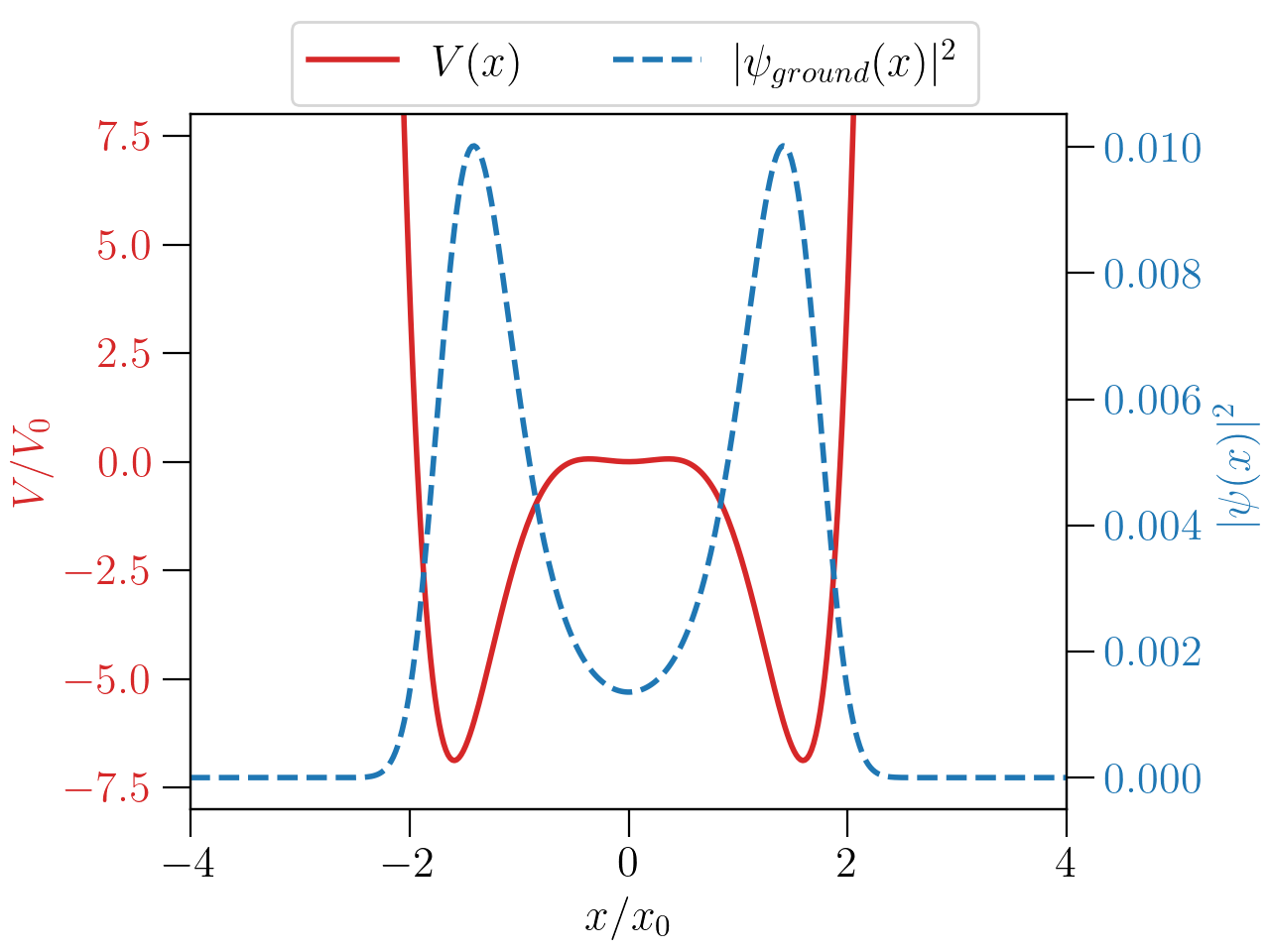}
    \caption{The polynomial triple-well potential (solid red line) given by $V(x) = x^2 - 4x^4 + x^6$ in units of $V_0=\frac{\hbar^2}{2Md^2}$ and $x_0=d$, where $d$ roughly corresponds to half the distance between the two wells of the potential and its ground state wave function (dashed blue line). Assuming a mass of $M=10^{-11} \; \si{\kg}$ and $d=10^{-5} \; \si{\meter}$, the ground-state energy is $-1 \times 10^{-47} \; \si{\joule}$, and the first excited energy is $-8.86 \times 10^{-48} \; \si{\joule}$. The potential gradient at $x=x_0$ is $4 \times 10^{-42} \; \si{\joule / \meter}$. For this particular potential, the potential gradient might be too shallow to be realizable.}
    \label{fig:potentials}
\end{figure}

In the previous section we assumed that the Zeno freezing of an arbitrary state is possible, without specifying what that state is and how one would achieve that. The first experimental demonstrations of the QZE froze internal energy levels of atoms \cite{Itano1990,Itano_2009}; similar studies followed for superconducting qubits \cite{Siddiqi16,siddiqi18} and NV centers in diamond \cite{Childress06,Wolters13}. When it comes to motional states, QZE has been used to suppress the escape of atoms from trapping potentials \cite{gagen93,altenmuller94,Fischer01}, and also in the context of many-body systems \cite{Patil15,Elliot16}. As for positional states, there have been proposals to generate a positional Schrödinger-cat state for atoms in a lattice inside an optical cavity state using QZE \cite{mekhov2012quantum, elliott2015multipartite}; to the best of our knowledge, the experiment has not yet been performed. A full proposal goes beyond the scope of this paper, but we can gather some basic ideas by assuming some features of a possible implementation.

Concretely, we assume that \textit{the desired state $\ket{\phi}$ is the non-degenerate ground state for the potential, in which the source system is trapped}: $H_S\ket{\phi}=E_0\ket{\phi}$. In such an implementation, the system is cooled to the ground state, and the Zeno effect is implemented as a frequent (ideally continuous) monitoring of the energy. We notice that the recent cooling of one harmonic mode of a nanosphere \cite{Deli2020} was, indeed, achieved with continuous probing.

For definiteness, let's look at \textit{symmetric multiwell potentials} \footnote{Since harmonics traps already exist, a good candidate for a delocalized state would seem to be the first excited state of the harmonic oscillator, that features two peaks separated by $d\approx \sqrt{\hbar/M\omega}$. But for the masses that we are interested in, given a reasonable trapping frequency $\omega$, $d$ is orders of magnitude smaller than the source itself. Indeed, take for the source a sphere of radius $R$ and constant density $\rho$, i.e.~$M=\frac{4\pi}{3}\rho R^3$. Already for the parameters of Ref.~\cite{Deli2020}, where the sphere is too small to show gravitational effects, we obtain a value of $d=10^{-11} \; \si{\meter}$ for $R=10^{-7} \; \si{\meter}$, $\rho=2600\;\si{\kg / \meter^3}$ and $\omega=10^5 \; \si{\hertz}$. The discrepancy will increase further for higher masses.} in one dimension (of course, in a full-fledged proposal one would have to achieve freezing in all three dimensions, as well as in the rotational degrees of freedom). The ground state of the double-well potential is almost-doubly-degenerate with tunneling between the two wells \cite{jelic2012double} and hence is unsuitable for our purposes; some symmetric triple-well potentials, however, have a nondegenerate delocalized ground state \cite{aquino2011energy} that might be usable for us. An example is plotted in Fig. \ref{fig:potentials}, a polynomial triple-well potential in the form 
\begin{align*}
    V(x) = ax^2 - bx^4 + cx^6.
\end{align*}

The shape of the potential must be very finely tuned to create the type of state that we want. For example, if the central well is too shallow, the whole potential is effectively a double well; if it is too deep, the ground state is essentially the ground state of the harmonic oscillator. Optimizing these parameters or choosing other forms for the potential, will be left for further studies.

While we will not propose an explicit scheme for performing the Zeno measurement in this paper, we envision it to be done through measurement of auxiliary fields which are conveniently already present in existing proposals for preparing genuinely quantum states. For example, in cavity cooling, a cavity laser is used to continuously drive the target into the ground state \cite{Windey2019,Delic2019}; in dissipative preparation of nonclassical states, the target is coupled to an auxiliary mode which dissipates into a low-temperature reservoir \cite{Filho96,Brunelli19}. Since the eventual Zeno scheme would most likely involve cooling and/or driving the source into the desired state, one possibility for us would be to perform the Zeno freezing by continuously monitoring these auxiliary fields. Of course, an analysis of how these fields affect the scattering pattern needs to be done.

\section{The scattering of the probe}
\label{sec:proposal}

\subsection{Setup}

\begin{figure}
    \centering
    
    \vspace{1 cm}
    
    \includegraphics[width=0.75\linewidth]{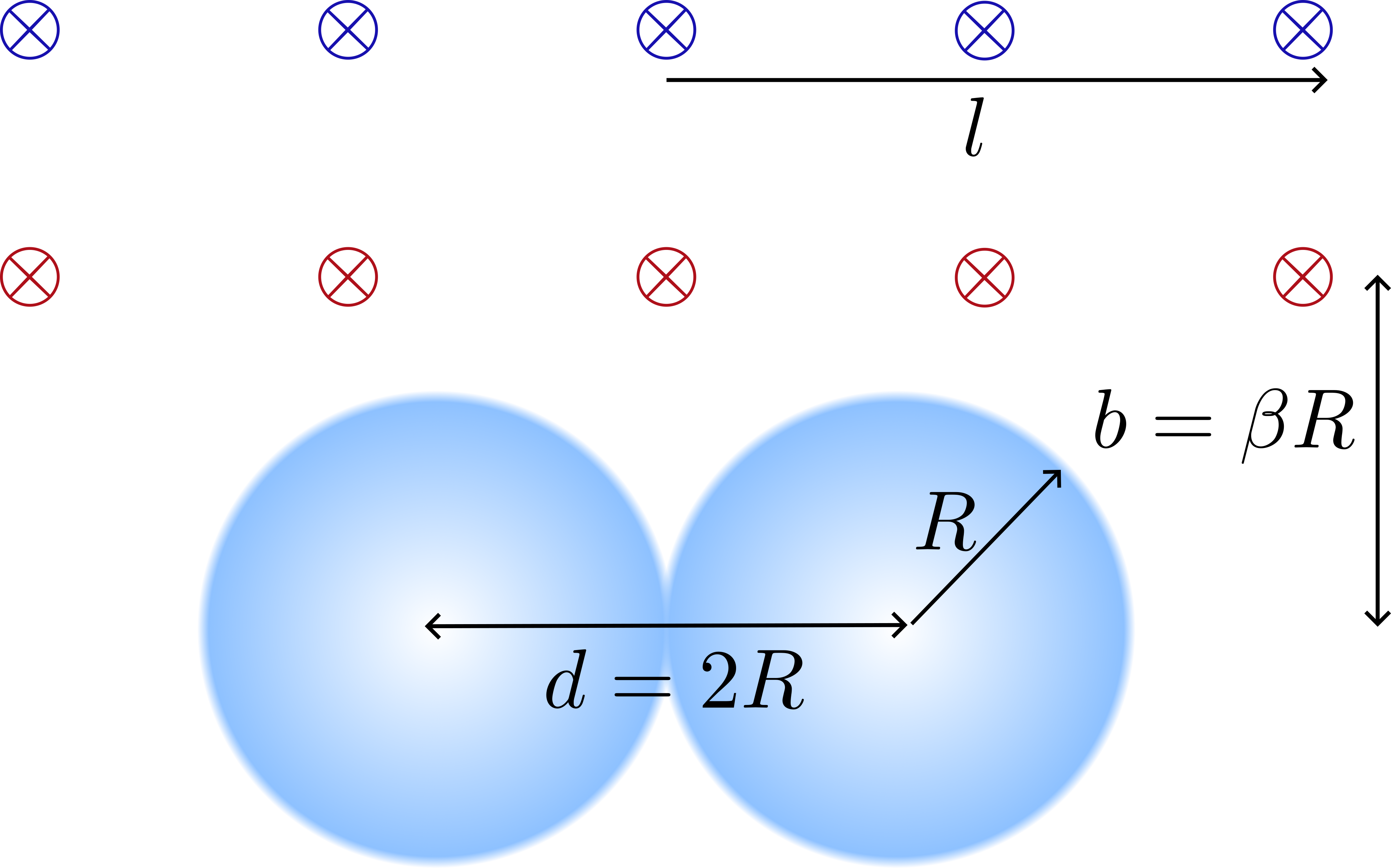}
    
    \vspace{1 cm}
    
    \includegraphics[width=0.75\linewidth]{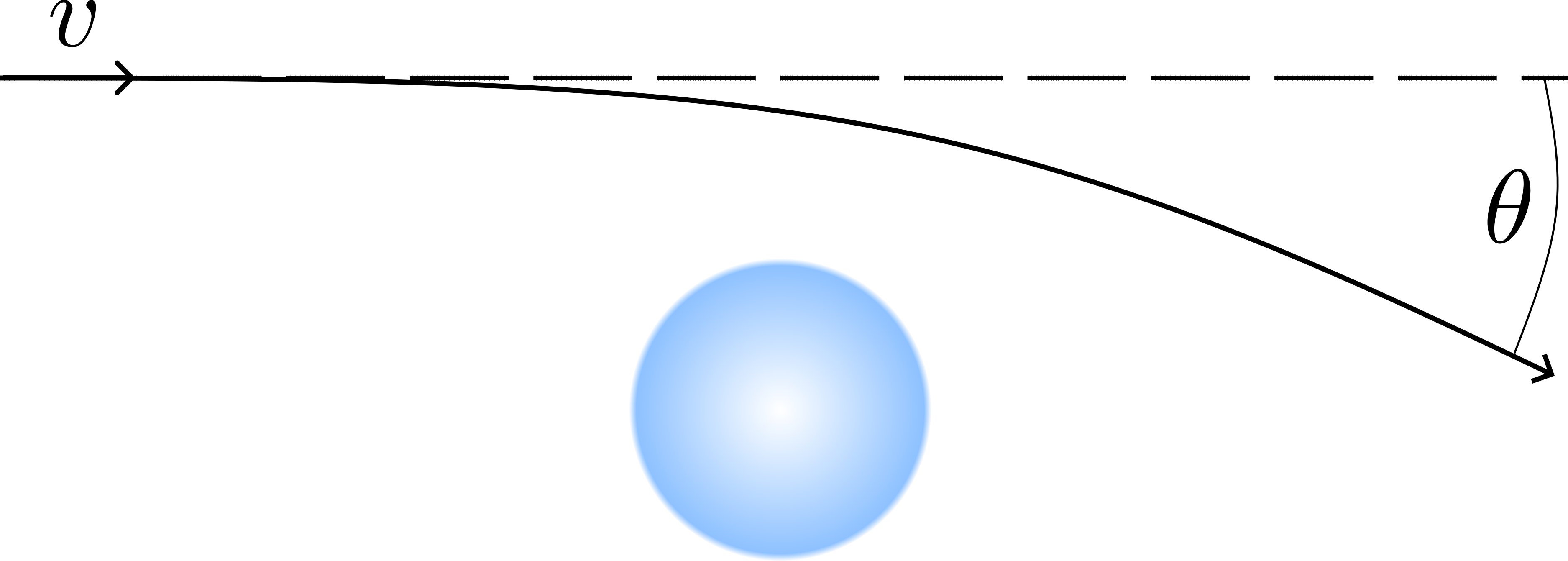}
    \caption{The setup for the experiment. \textbf{Top:} The two blue spheres represent not two different source objects, but one source object in a superposition of being in two different positions. The red and blue arrows represent the initial coordinates of incoming probe particles, with different colors signifying different impact parameters $b$. \textbf{Bottom:} The scattering or deflection angle $\theta$ is the angle between the initial velocity and the outgoing velocity.}
    \label{fig:diagram}
\end{figure}

For our feasibility study, we consider the basic case: the wavefunction of the source is delocalized along the $x$ axis, symmetric around the origin, and localized along the other axes. The scattering direction is the $z$ axis. The probes are sent from $(l, b, -\infty)$ with velocity $(0,0, v)$, where we call $b$ the scattering parameter and $l$ the offset. A graphical representation of the setup can be seen in Fig. \ref{fig:diagram}. For simplicity, we consider scattering parameters such that the probe does not hit the source's wavefunction.



A major concern in detecting gravitational effects is to avoid other interactions that are typically much stronger, like the electromagnetic interactions. Choosing electrically neutral objects for either the source or the probe is an obvious requirement, but whether it is sufficient or not depends on the actual realization. For instance, if the source is optically levitated, it will develop an induced dipole, and so will the probe, since it will have to fly through the tweezer field to come close to the probe. Probing an electrically levitated source with a neutral probe may be a better scheme to have in mind, although at this stage we are not committing to a specific platform.

\begin{figure*}
    \centering
    \includegraphics[width=0.9\linewidth]{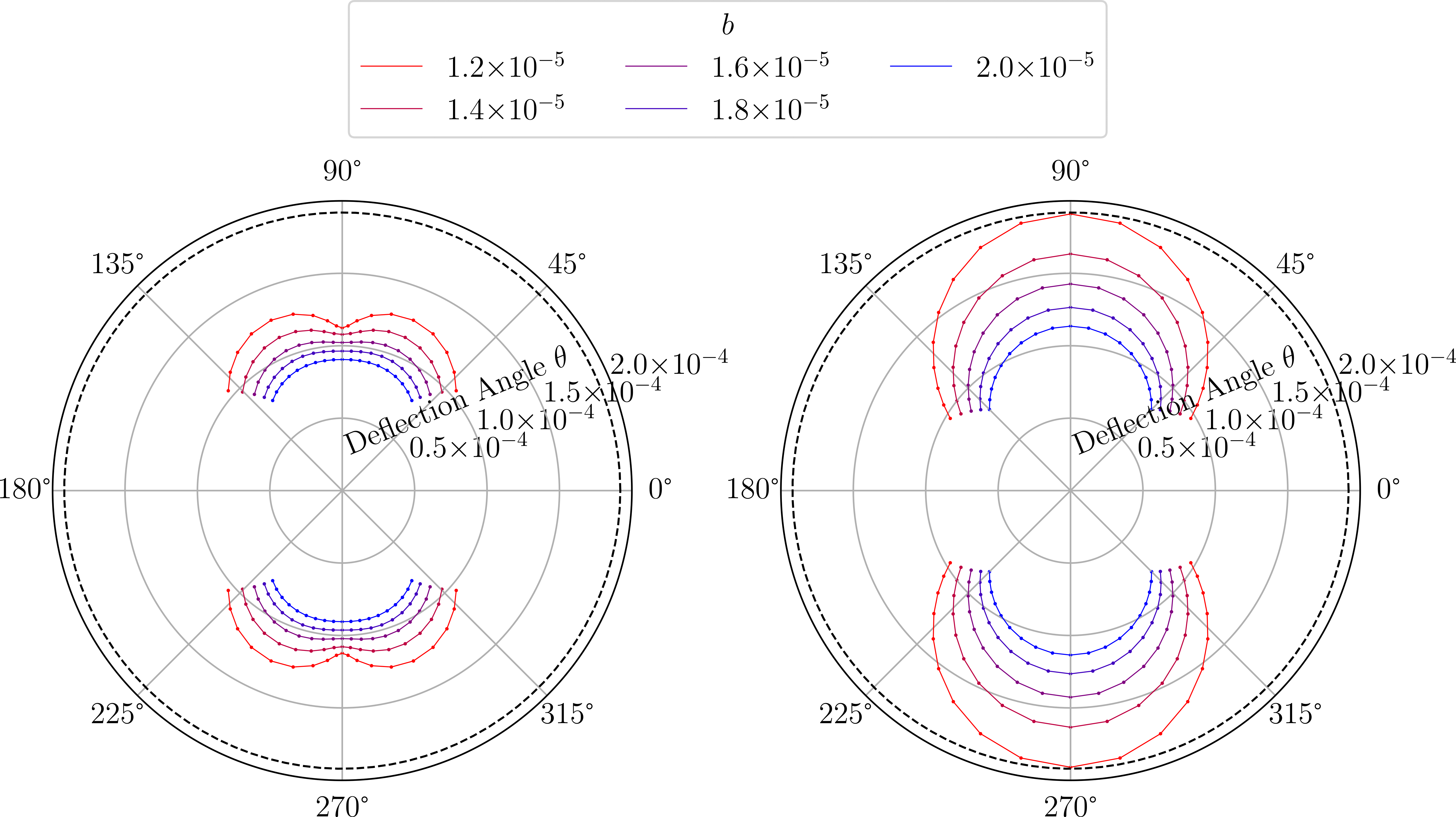}
    \caption{A stereographic projection of the outgoing angles of the probe from the negative $z$ axis. \textbf{Left:} The setup described in Fig. \ref{fig:diagram} and the parameters $R\equiv d/2 =10 \; \si{\micro \meter}$, $\rho = 2600 \; \si{\kg / \meter^3}$, $t_R = 10^{1.1}$, $\beta\in [1.2 , 2.0]$, and $l \in [0 , 2R]$. \textbf{Right:} For comparison, the same setup, but now the source is just a single, fully classical, local, and stationary sphere localized at the origin. The dashed lines in both plots correspond to Eq. \eqref{eq:max_deflection}. This result is obtained by running a numerical simulation.}
    \label{fig:cross_section}
\end{figure*}

\subsection{Detecting the signal}

The outgoing angles of the probes are determined by the shape of the source object. We ran numerical simulations of the scattering using the setup that we have just described with the source wavefunction being $\phi(\mathbf{X}) = 1/2 [\delta(X+d/2) + \delta(X-d/2)]$. The simulation is a standard Newtonian scattering simulation with a mass distribution $\rho(\mathbf{X}) = M|\phi(\mathbf{X})|^2$. For simplicity, the source is completely frozen, corresponding to an infinitely fast Zeno measurement. The difference in the scattering pattern between the delocalized source and that of a sphere localized at the origin is presented in Fig.~\ref{fig:cross_section}. The values of the parameters needed to see this difference in an experiment are estimated in the next section.

The collapsed situation, in which the sphere gets localized either at $-d/2$ or at $-d/2$ for each probe, is even easier to distinguish: indeed, for given impact parameters, the probes would follow either of two trajectories. If on the contrary one observed a single trajectory for every impact parameter, the scattering potential would be the same for each probe.

Revisiting here what was anticipated in Sec. \ref{sec:glance}, the same scattering pattern as the delocalized Zeno-frozen source can be obtained from a classical source only in two ways. The first is by creating the static distribution $|\phi(\mathbf{X})|^2$, but this means that the sphere morphs itself into two smaller spheres during the experiment and returns back to its original shape thereafter, which is absurd. The second is by undergoing a fast motion along the $x$ axis, such that its position within the scattering time appears to be distributed according to $|\phi(\mathbf{X})|^2$. However, given a potential in which the source is trapped, the natural classical trajectories are the Newtonian ones in that potential, and it is unlikely that fast motion along one of those trajectories matches our desired distribution. Thus, if we observe the scattering pattern corresponding to a delocalized sphere, we can conclude that the delocalization of the source is of quantum origin.

\section{Feasibility}
\label{sec:feasibility}

\begin{figure}
    \centering
    \includegraphics[width=\linewidth]{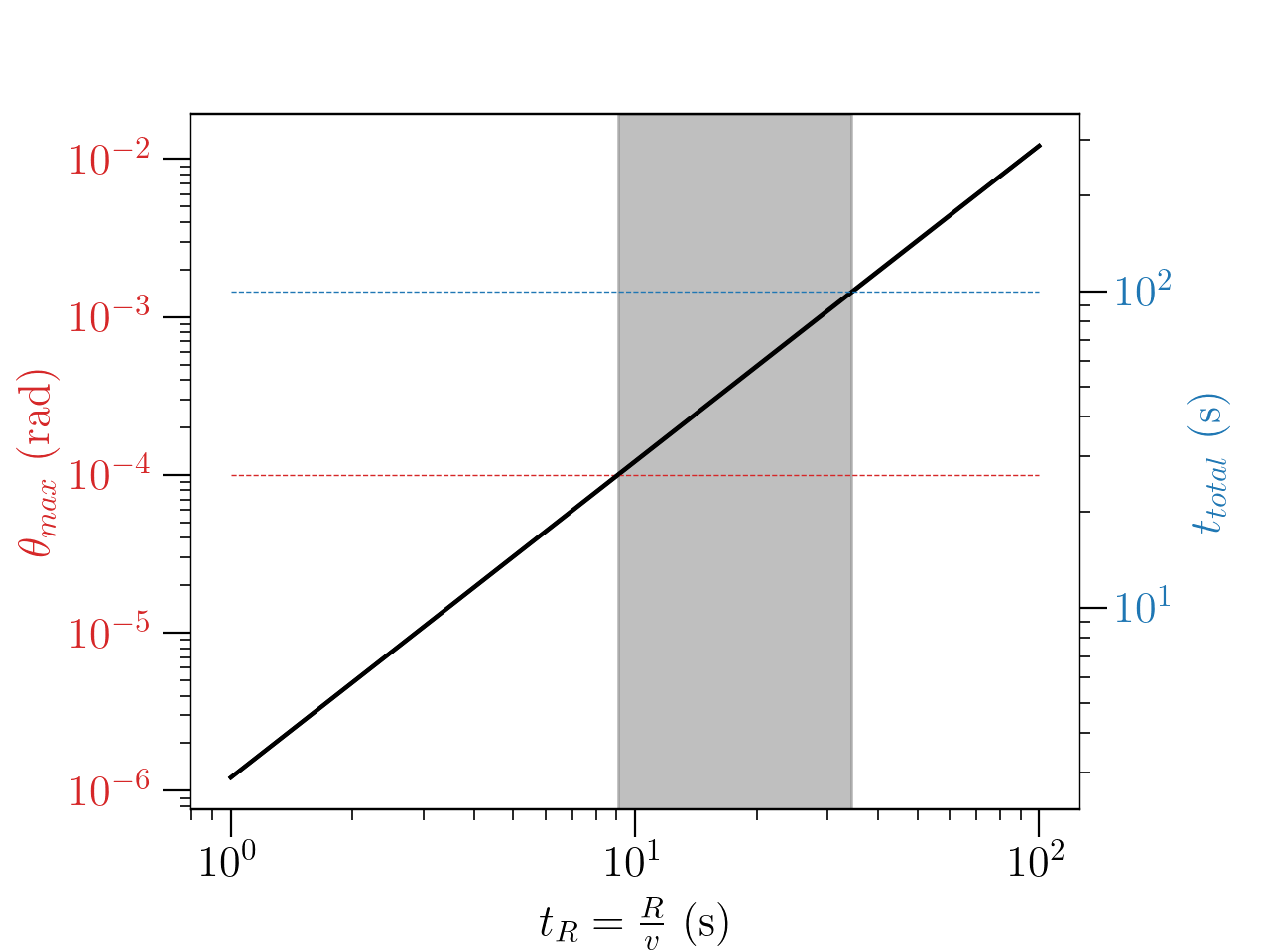}
    \includegraphics[width=\linewidth]{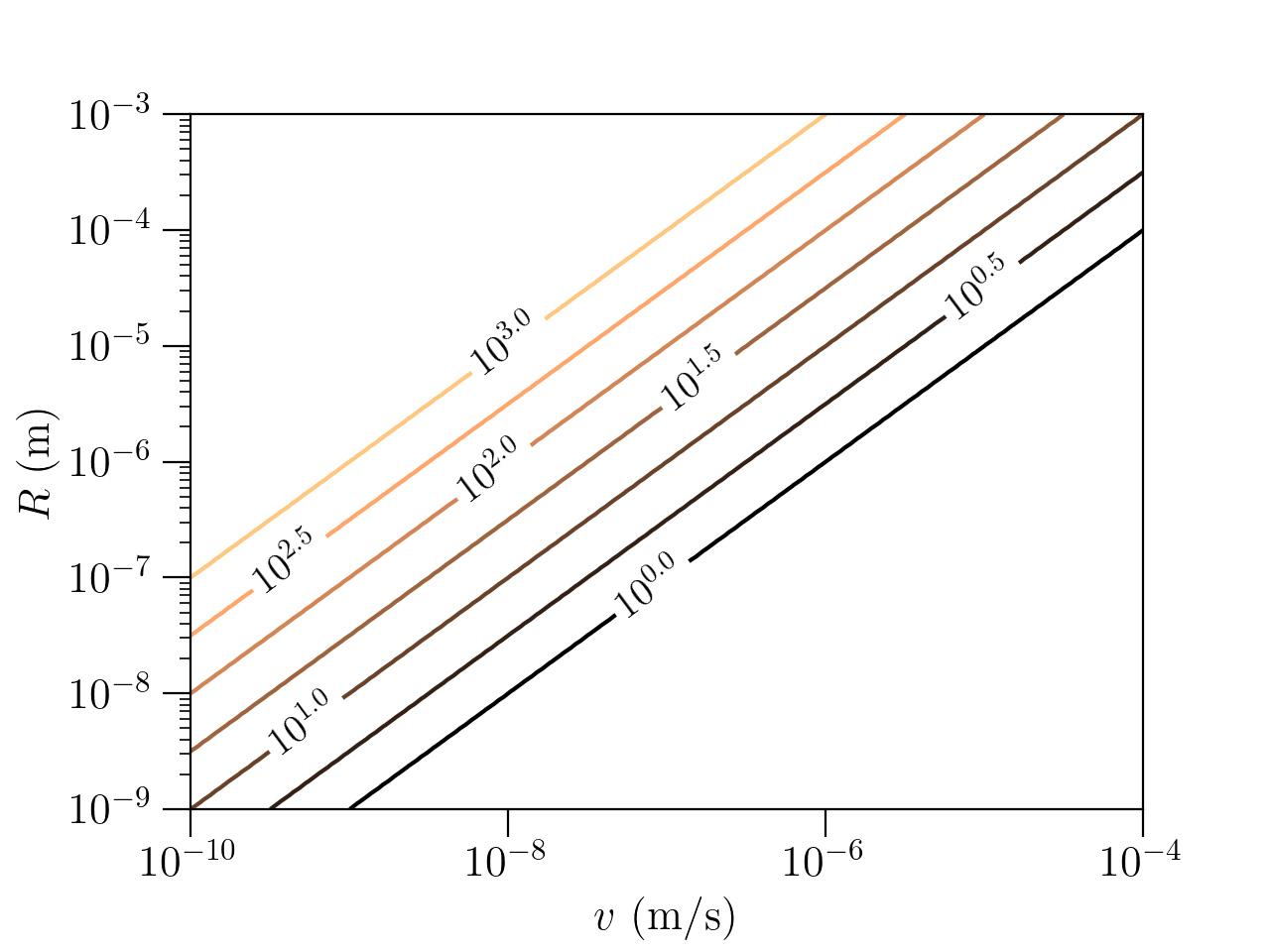}
    \caption{\textbf{Top}: The relationship between the maximum deflection $\theta_{max}$ (left $y$ axis) and the total scattering time $t_{total}$ (right $y$ axis) with the parameter $t_R$ ($x$ axis) from Sec. \ref{subsec:angle and time} (with $\beta = 1.2$ and $\zeta = 0.75$). The gray region shows the values of $t_R$ that are most realiztic.  \textbf{Bottom}: Contour plots of $t_R=R/v$ in the $(R,v)$ plane.}
    \label{fig:scattering_parameters}
\end{figure}

In this section we study the feasibility of such a proposal. The numerical values that we choose are indicative. For instance, the density of the sphere is chosen as that of silica, $\rho=2600 \; \si{\kg / \meter^3}$. We will show that there is a feasibility region, and we identify the most constraining factor.

\subsection{Deflection angle and scattering time}
\label{subsec:angle and time}

The first constraints that we consider come from considering the maximum deflection angle of the probe $\theta_{max}$ and the time it takes for the scattering process $t_{total}$. We want a value of $\theta_{max}$ that is larger than our detector's resolution and a value of $t_{total}$ that is sensible. This will result in a bound on the time
\begin{equation}t_R = \frac{R}{v}\end{equation} needed by a probe with velocity $v$ to cover the distance $R$.

For the estimates in this section, rather than having the source in superposition, we consider scattering by a spherical source of radius $R$ in a classical stationary state, and we lift the formulas from Rutherford scattering, with gravity instead of electromagnetic forces.

The largest deflection angle $\theta_{max}$ will be obtained by sending the probe as close as possible to the source while not hitting it. We write the minimum impact factor as $b_0 = \beta R$, with $\beta > 1$. In terms of these parameters, the deflection angle is given by \cite{taylor2005classical}
\begin{align}\label{eq:max_deflection}
    \theta_{max} &= 2 \cot^{-1}\left(\frac{v^2 b_0}{GM}\right)\,\approx\, \frac{8\pi G \rho}{3\beta} t_R^2\,.
\end{align}

The scattering time needs to be defined since the trajectory of the probe is a hyperbola. We define it as twice the time it takes for the probe to fly from the periapsis (the closest approach) to the point where the true anomaly $\phi$ (the angle between the periapsis and the position of the probe) is a fraction $\zeta$ of the asymptotic true anomaly $\phi_\infty$. The latter is related to the scattering angle $\theta$ by the equation $\theta + 2\phi_\infty + \pi = 0$, and the time taken for a particle to fly from the periapsis to a point with an anomaly $\phi$ is given by \cite{curtis2013orbital}
\begin{align*}
    \frac{(GM)^2}{h^3}t &= \frac{1}{e^2-1} \frac{e\sin\phi}{1+e\cos\phi} \\&- \frac{1}{(e^2-1)^{3/2}} \ln\left(\frac{\sqrt{e+1}+\sqrt{e-1}\tan\frac{\phi}{2}}{\sqrt{e+1}-\sqrt{e-1}\tan\frac{\phi}{2}}\right)\,.
\end{align*}
In this equation, $h$ is the angular momentum per unit mass of the particle, defined through ${(GM)^2}/h^3 = \left(\frac{4}{3}\pi G \rho \right)^2(t_R^3/\beta)^3$, and $e$ is the eccentricity defined through $\cos^{-1}\left(-1/e\right)=\phi_\infty$.

In summary, besides $\rho$ and our safety margin parameters $\beta$ and $\zeta$, $\theta_{max}$ and $t_{total}$ depend only on $t_R$. The reasonable bounds $\theta_{max} > 10^{-4} \; \si{\radian}$ and $t_{total} < 10^2 \; \si{\second}$ yield the rather tight range $t_R\in[10^1,10^{1.2}]\,\si{\second}$ for $\beta = 1.2$ and $\zeta = 0.75$ (Fig.~\ref{fig:scattering_parameters}, top panel). For instance, if we choose a sphere with $R\approx 10 \; \si{\micro\meter}$, the probe's velocity will have to be $v\approx 1 \; \si{\micro\meter / \second}$ (Fig.~\ref{fig:scattering_parameters}, bottom panel).

\subsection{Decoherence}
\label{subsec:decoherence}

Next, we discuss the localization of the source by decoherence, the obstacle that we aim to defeat by employing the Zeno effect. Here, we will follow Joos and Zeh's treatment of decoherence \cite{joos2013decoherence, RomeroIsart2011, aspelmeyer2022avoid}. The evolution equation is
\begin{equation}
    \bra{x}\Dot{\rho}(t)\ket{x'} = \frac{i}{\hbar} \bra{x} [\hat{\rho}, H] \ket{x'} - \Gamma_D(x-x')\bra{x}\rho(t)\ket{x'}.\label{zeh}
\end{equation}
with the rate of decoherence
\begin{equation}
    \Gamma_D(x) = \lambda_{th}^2\Lambda \left(1 - \exp \left(-\frac{x^2}{\lambda_{th}^2}\right) \right),
\end{equation}
where $\Lambda$ is called localization parameter and $\lambda_{th}$ is the thermal wavelength of the scattering particles. As an estimate, we are concerned with $\Gamma_D(R)$, the decoherence rate at a distance $R$.

In our study we take into account two unavoidable sources of decoherence: the interaction with the rest gas and blackbody radiation. Next, we distill the useful information from \cite{aspelmeyer2022avoid, RomeroIsart2011}. We do not address the decoherence due to displacement and frequency noise in the trapping potential \cite{Schneider99,Gehm98}, as it will depend on the exact set-up used to prepare the Zeno-frozen source, which we have decided to leave for future work.

\subsubsection{Rest gas}

Since a perfect vacuum does not exist, the air molecules localize the source by colliding with it. This process is closely connected to the pressure of the environment. The thermal wavelength for this process is given by
\begin{equation}
    \lambda_{th} = \frac{2\pi\hbar}{\sqrt{2\pi m_{H_2}k_B T_e}},
\end{equation}
where $T_e$ is the temperature of the environment and $m_{H_2}$ is the mass of a hydrogen molecule (taken to be the rest gas). Because the thermal wavelength is usually much smaller than the superposition size, we could approximate $\Gamma_g \approx (\lambda_{th})^2\Lambda$. With
$$
\Lambda = \frac{8\sqrt{2\pi} m_{H_2} \Bar{v} p R^2}{3\sqrt{3} \hbar^2},
$$
where $\Bar{v}$ is the mean velocity of the air molecules, the decoherence by the rest gas is given by
\begin{equation}
    \Gamma_g = \frac{\lambda_{th}}{\hbar} \frac{16\pi}{3} p\,R^2,
\end{equation}
where $R$ is the radius of the particle and $p$ is the pressure. Inserting the constants, $\Gamma_g \approx 1.96 \times 10^{26} \frac{pR^2}{\sqrt{T_e}}$.

\subsubsection{Blackbody radiation}

There are three decoherence channels associated with thermal photons, namely scattering, absorption, and emission. The thermal wavelength of decoherence by blackbody radiation is given by $\lambda^{e,i}_{th} = \frac{\pi^{2/3}\hbar c}{k_B T_{e,i}}$, where $T_{e,i}$ is the temperature of the environment or the internal temperature, depending on the decoherence considered. The thermal wavelength is usually larger than the superposition size. Thus, we could use the approximation $\Gamma(x) \approx \Lambda x^2$. The rate of decoherence is given by the $\Lambda$ parameters:
\begin{enumerate}
    \item For scattering,
    \begin{equation}
        \Lambda_{bb}^{sc} = \left( \frac{1}{\lambda^e_{th}} \right)^9 \left(\frac{8!8 \zeta (9) \pi^5 c R^6}{9} \right) \text{Re}\bigg( \frac{\epsilon -1}{\epsilon+2}\bigg)^2
    \end{equation}
    \item For absorption and emission,
    \begin{equation}
        \Lambda_{bb}^{ab,em} = \left( \frac{1}{\lambda^{e,i}_{th}} \right)^6 \left(\frac{16 \pi^9 c R^3}{189} \right) \text{Im}\left( \frac{\epsilon -1}{\epsilon+2}\right)
    \end{equation}
\end{enumerate}
where $\epsilon$ is the dielectric constant of the material. If we assume the worst-case scenario of large dispersion and absorption, $\frac{\epsilon-1}{\epsilon+2} \approx 1$, we get

\begin{align*}
    \Lambda_{bb}^{sc} \approx 5 \times 10^{36} R^6 T_e^9, \\
    \Lambda_{bb}^{e,i} \approx 5 \times 10^{25} R^3 T_{e,i}^6.
\end{align*}

\subsubsection{Bounds on the Zeno rate}

\begin{figure}
    \centering
    \includegraphics[width=\linewidth]{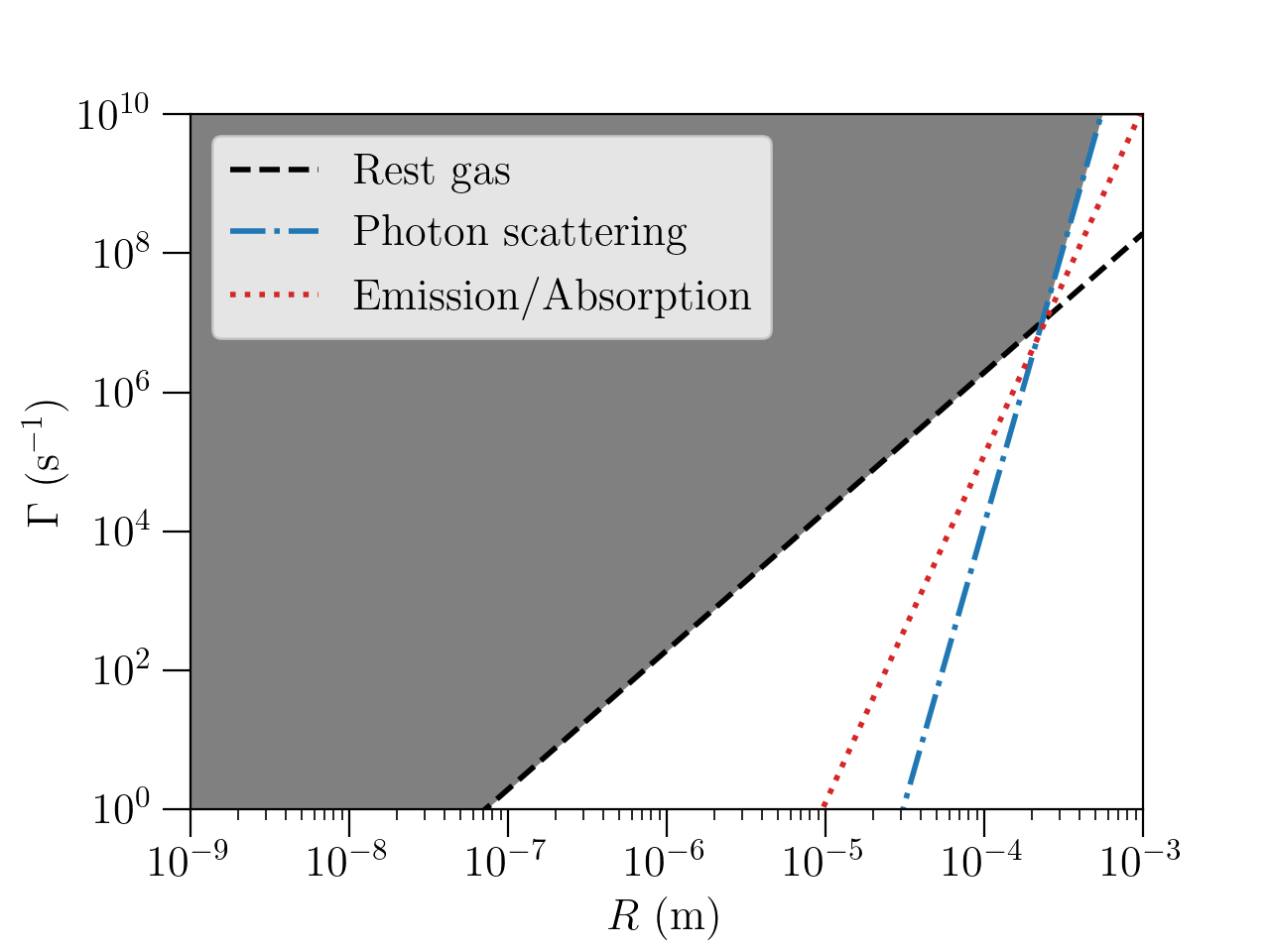}
    \includegraphics[width=\linewidth]{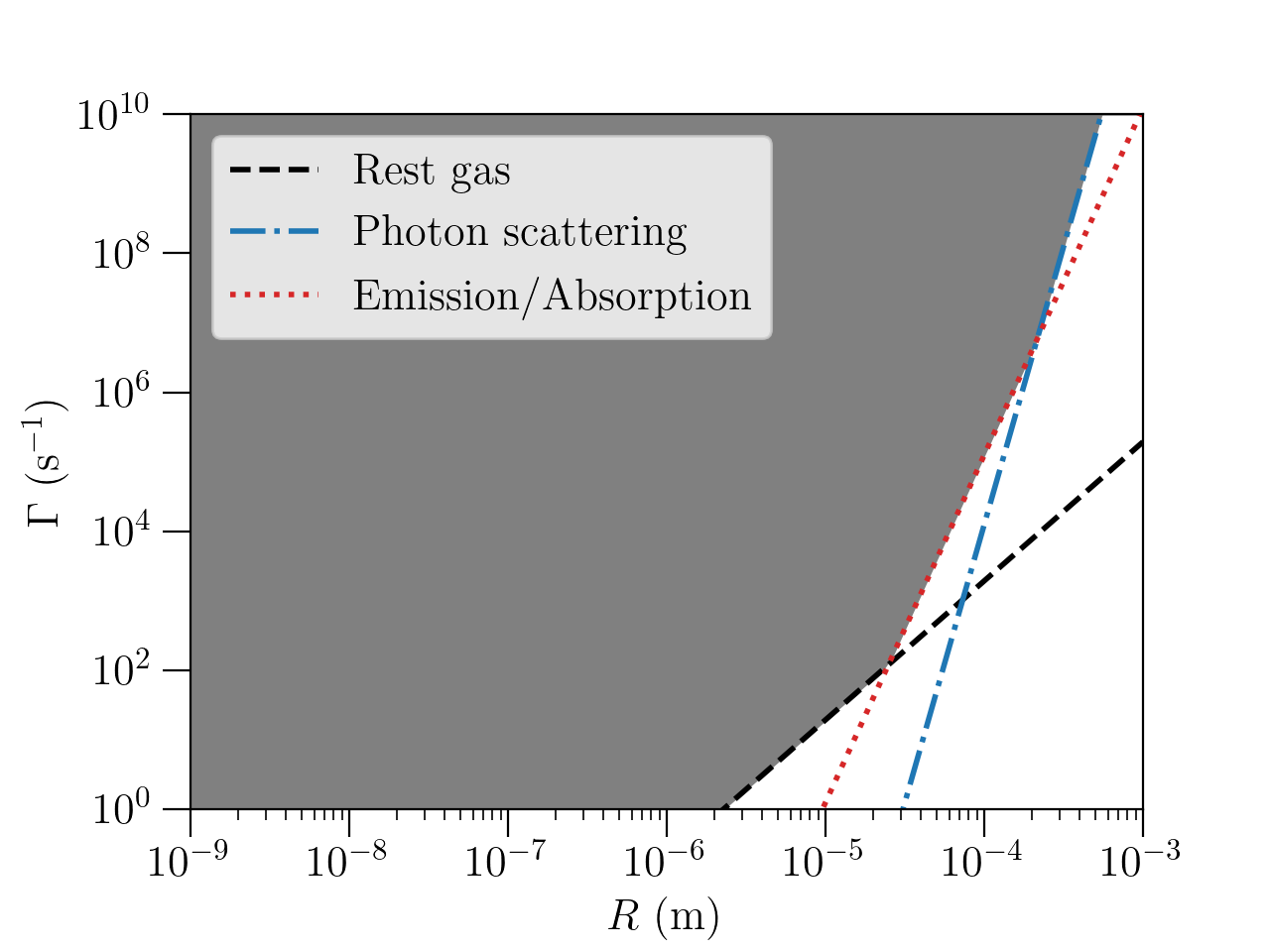}
    \caption{Plots of the lower bounds of the Zeno rate $\Gamma_{Zeno}$ from decoherence (from Sec. \ref{subsec:decoherence}) for a pressure of $10^{-12}\; \si{\pascal}$ (\textbf{top}) or $10^{-15}\; \si{\pascal}$ (\textbf{bottom}). The gray region is the allowed region for $\Gamma_{Zeno}$. $T=1 \; \si{\kelvin}$ for all cases. As can be seen, in different regimes the main sources of decoherence are different.}
    \label{fig:zeno_region}
\end{figure}

Without going into a detailed description of Zeno dynamics under decoherence, we can approximate that, in order to keep the source in a coherent superposition, we would need the Zeno measurement to be applied much faster than all the decoherence rates,
\begin{equation}
    \Gamma_{Zeno} \gg \Gamma_{D}(R),
\end{equation}
where $\Gamma_{Zeno}:=1/\tau$.

The values of $\Gamma_{D}(R)$ for the various decoherence channels are plotted in Fig.~\ref{fig:zeno_region}, assuming that the system is cooled to $T_e = T_i = 1 \; \si{\kelvin}$ and that pressures down to $10^{-15} \; \si{\pascal}$ are achievable. In this case, a superposition of a sphere of a radius of up to $10^{-5}\;\si{\m}$ could be maintained using reasonable Zeno rate of $10^2$ to $10^3\;\si{\second^{-1}}$. 


While these give an upper bound of the required rate of Zeno measurements, there is another constraint that we must account for, namely, the Zeno dynamics itself.

\subsection{Zeno dynamics}
\label{subsec:zeno dynamics constraints}

\begin{figure}
    \centering
    \includegraphics[width=\linewidth]{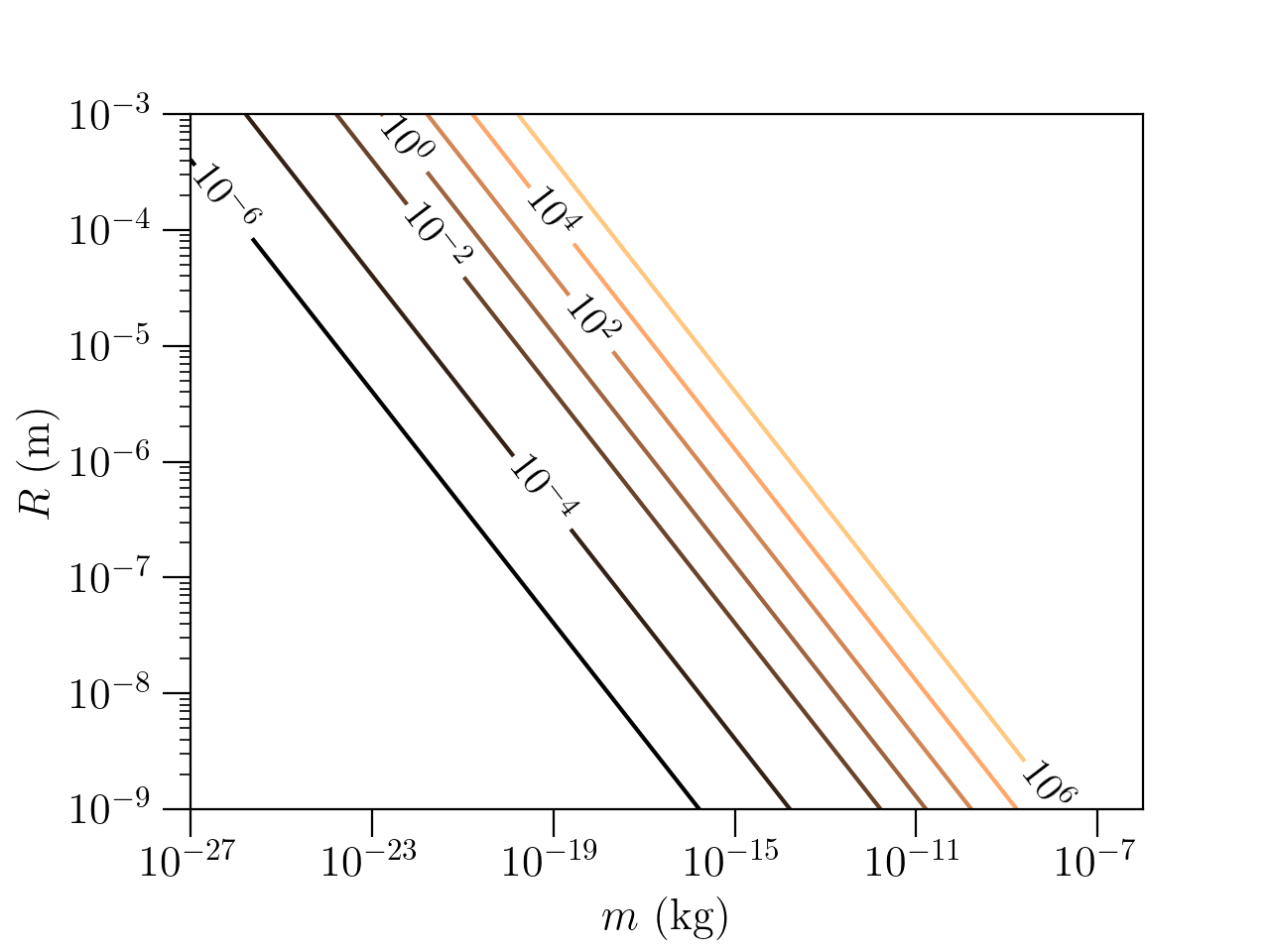}
    \caption{Lower bounds of the Zeno rate $\Gamma_{Zeno}$ from Zeno dynamics (from Sec. \ref{subsec:zeno dynamics constraints}). The plotted function is $\max(\tau^{-1}_{Z}, 100\tau^{-2}_{Z})$ from Eqs. \eqref{eq:zeno1inv} and \eqref{eq:zeno2inv} with $t_{total}=100 \; \si{\second}$.}
    \label{fig:zeno_dynamics}
\end{figure}

As we have discussed in Sec. \ref{subsec:zeno}, the dynamics of Zeno freezing itself also provides some constraints on the rate of Zeno measurements. Referring back to Eq. \eqref{eq:zeno1} we have to have $\tau \ll \tau_{Zeno}$, or, equivalently,
\begin{equation}\label{eq:zeno1inv}
    \Gamma_{Zeno} \gg \tau^{-1}_{Z} \approx \hbar^{-1}\,\left(\frac{GmM}{b_0}\right).
\end{equation}

Second, due to the constraint that the system should have a reasonable probability to survive until the end of the experiment, referring back to Eq. \eqref{eq:zeno2}, the Zeno rate gives an upper bound on the duration of the experiment. However, since we (arbitrarily) decided that $t_{total}$ should not exceed $100 \; \si{\second}$ in Sec. \ref{subsec:angle and time}, we turn this into another constraint for $\Gamma_{Zeno}$,
\begin{equation}\label{eq:zeno2inv}
    \Gamma_{Zeno} \gg \frac{t_{total}}{\tau_Z^2}.
\end{equation}
This constraint is plotted in Fig. \ref{fig:zeno_dynamics}. Together with Fig. \ref{fig:zeno_region}, this gives a lower bound on $\Gamma_{Zeno}$.

\subsection{Other constraints on the probe}

\subsubsection{Consistency with the assumption of classical scattering}\label{subsec:classicality}

\begin{figure}
    \centering
    \includegraphics[width=\linewidth]{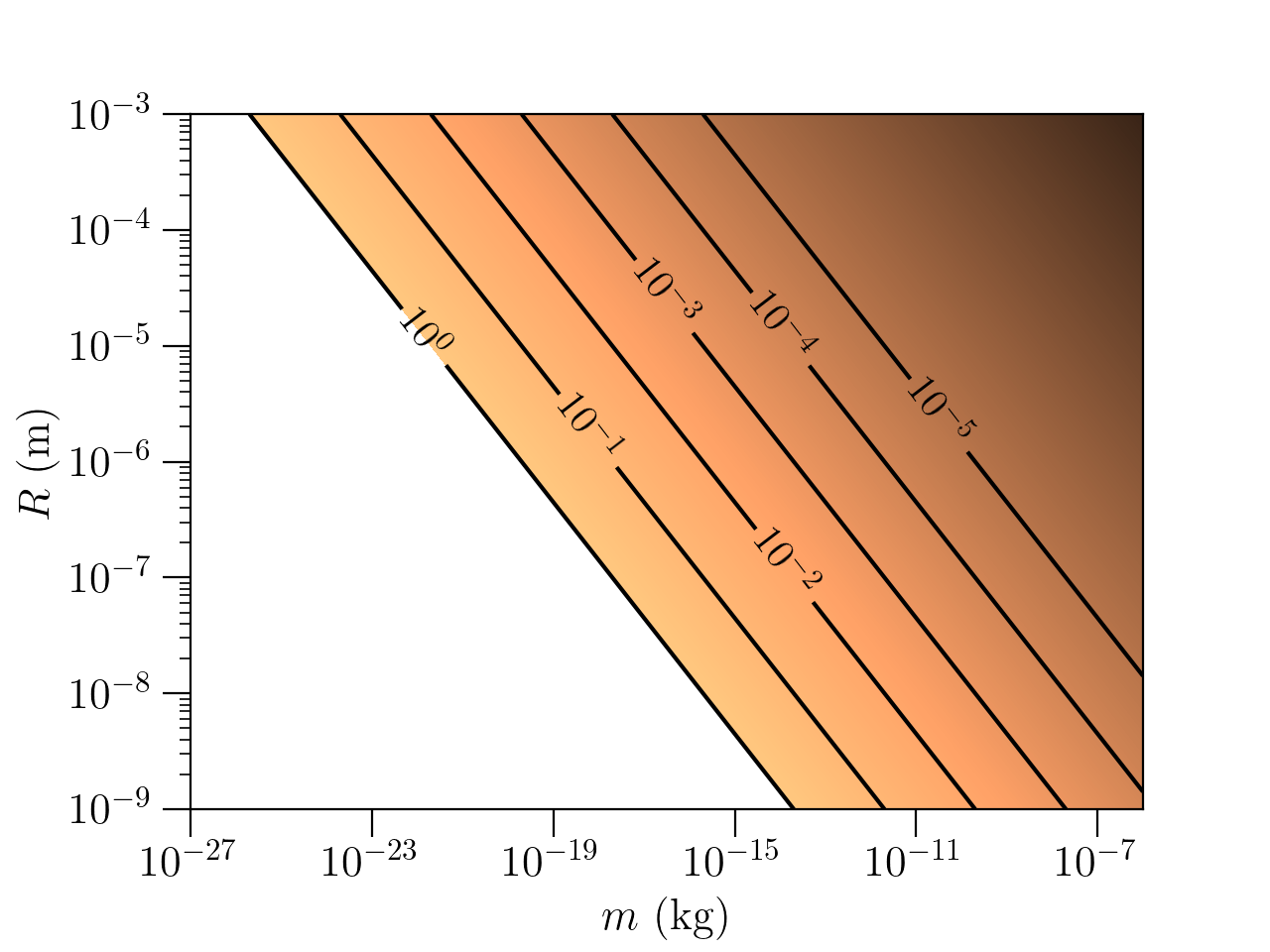}
    \caption{The assumption of classical scattering puts a lower bound on the mass of the probe. The shaded region is the allowed region of the probe's mass ($x$ axis) for a given radius of the source ($y$ axis); (from Sec. \ref{subsec:classicality}). The contour lines give the values of the ratio $\sigma_{min}/{R}$.}
    \label{fig:probe}
\end{figure}


In this paper, we have computed scattering with a classical trajectory, leaving the scattering of matter waves for future work. For this assumption to be valid, the probe's wave function must be sufficiently localized during the whole experiment. To study this constraint, since the interaction is weak and not confining, we borrow the spread of a Gaussian wavepacket $\sigma_u$ from the theory of the free particle: 
\begin{equation}
    \sigma_u(t) \approx \sqrt{2(\Delta u)^2 + \frac{1}{2}\left(\frac{\hbar t}{m \Delta u}\right)^2},
\end{equation}
where $u=x,y,z$ and $\Delta u$ is the uncertainty at $t=0$. The left0hand side reaches its minimum value $\sigma_{min} = \sqrt{2\hbar t/m}$ for $\Delta u_{min} = \sqrt{\hbar t/2 m}$. For the matter-wave behavior to be negligible, we require $\sigma_u(t_{total})\ll R$. The result of this analysis is plotted in Fig.~\ref{fig:probe} using $t_{total}=100 \; \si{\second}$: for a source with radius $R=10^{-5} \; \si{\meter}$, a probe with mass $10^{-18} \; \si{\kg}$ could be prepared such that its spread at the end is 1/100\textsuperscript{th} of $R$. This constraint constraints the same parameters as the one in Sec. \ref{subsec:zeno dynamics constraints}, leading to a trade-off between a lower required Zeno measurement rate and heavier and more localized probes.

As a final check, a wavepacket with $\Delta u = \Delta u_{min}$ will have $\Delta p_u \geq \sqrt{\hbar m/2 t_{total}}$. For the parameters that we are considering, this lower bound is much smaller than $p=mv$: thus, the momentum of the probe can also be prepared with a small spread.

\subsubsection{Mean free path of the probe}\label{subsec:mfp}

\begin{figure}
    \centering
    \includegraphics[width=\linewidth]{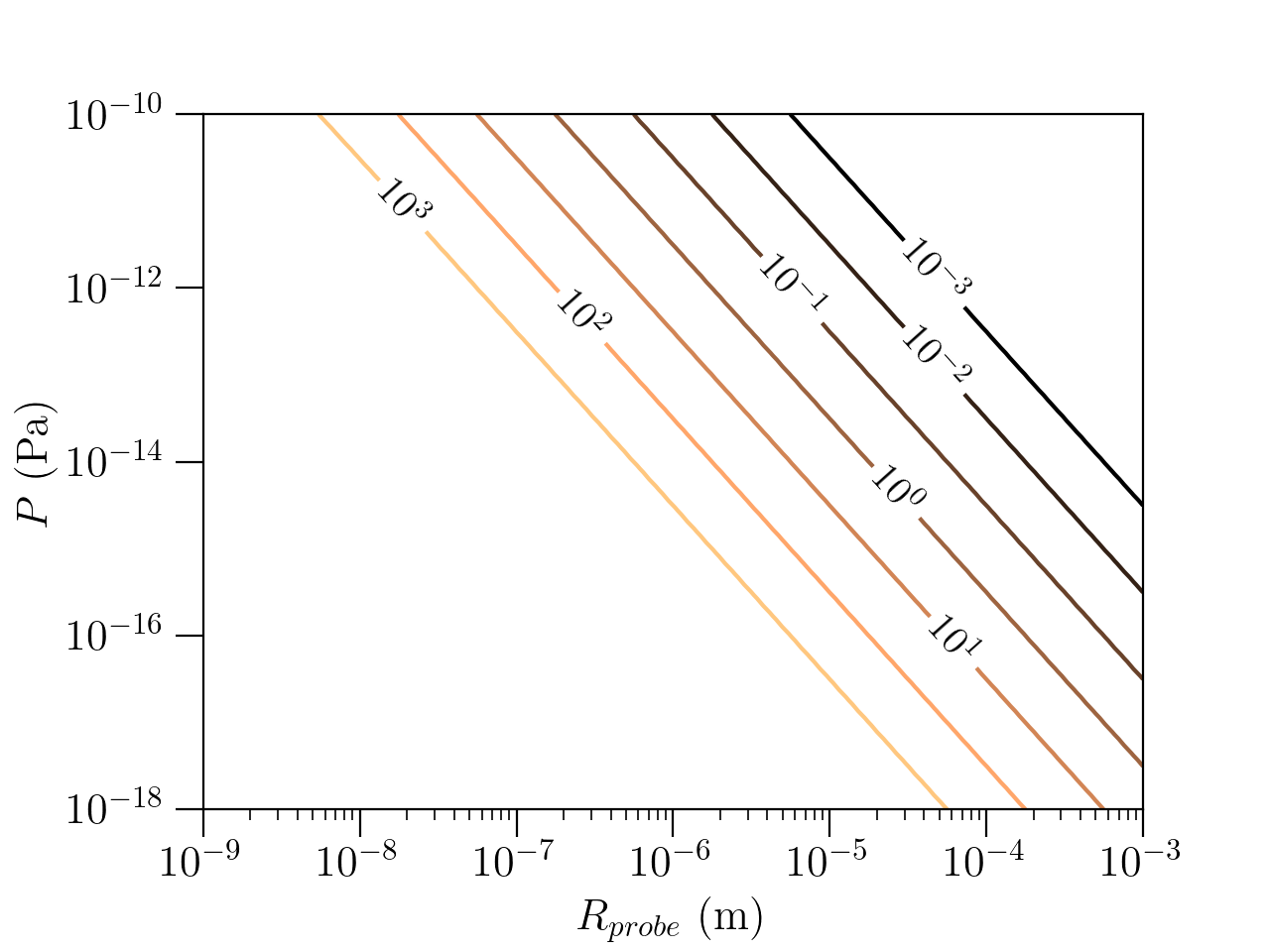}
    \caption{Plots of the mean free paths $l_{MFP}$ of different pressure and probe radius (from Sec. \ref{subsec:mfp}).}
    \label{fig:mean_free_path}
\end{figure}

Since the probe is slow, we should also check that the scattering trajectory is not blurred by interaction with the environment. The mean free path of the probe particle is

\begin{equation}
    l_{MFP} = \frac{k_B T_e}{\sqrt{2} A p} \approx 3.6 \times 10^{-15} \frac{T_e}{p},
\end{equation}
where $A=\pi\left( \frac{d_{H_2}}{2} + R_{probe}\right)^2$ is the collisional cross section of the sphere and an air molecule. Plotting for different values of pressure and probe size (Fig. \ref{fig:mean_free_path}), we find that for a sufficiently low pressure ($p<10^{-12} \; \si{\pascal}$) the mean free path would be very long ($\sim 10\; \si{\meter}$) for a probe with a size of up to $1 \; \si{\micro \meter}$. In conclusion, the deflection of the probe by collisions with the environment is not a major concern for this proposal.

\subsection{Summary: a feasibility region}

Let us summarize a possible feasibility region, using numbers that we have already used as examples above.

For the source, we start by setting $R = 10 \; \si{\micro \meter}$, corresponding to a mass of $10^{-11} \; \si{\kg}$ for a density of $2600 \; \si{\kg / \meter^3}$. This value for the mass of the source is similar to the estimate of other proposals \cite{Bose17, Marletto:2017kzi, aspelmeyer2022avoid} and is, indeed, midway in magnitude between the values cited in the Introduction for current detection of gravity and of quantum effects.

The main idea of our proposal is that decoherence can be fought by the Zeno effect. The bounds for decoherence (Sec. \ref{subsec:decoherence}) in a low-pressure cryogenic environment ($P=10^{-15} \; \si{\pascal}$, $T=1\;\si{\kelvin}$) imply that the Zeno measurement must be repeated with rate \begin{equation}
\Gamma_{Zeno} \gg 100 \; \si{\second^{-1}}\,.\label{zenonumber}\end{equation} This value provides a benchmark for future specific proposals aimed at preserving a coherent spatial superposition through the Zeno effect.

The detection phase of our proposal is done by classical scattering of a probe. To achieve sufficiently large scattering in a time on the order of a minute (Sec. \ref{subsec:angle and time}), we need $t_R=R/v\approx 10 \; \si{\second}$, i.e., a probe speed $v \approx 1 \; \si{\micro \meter / \second}$. Consistency with the assumption of classical scattering dynamics requires the probe's mass to be $m \approx 10^{-18} \; \si{\kg}$  (Fig.~\ref{fig:probe}). The kinetic energy of the probe will correspondingly be $3 \times 10^{-12} \; \si{\eV}$. The window of possible values for $t_R$ is one of the tightest constraints of this proposal, and forces the use of a low-energy probe.

With these values, other constraints prove less critical. Notably, the disturbance of the source by the probe turns out to be similar to that by decoherence, and can therefore be compensated by the same Zeno rate \eqref{zenonumber} (Sec. \ref{subsec:zeno dynamics constraints}); and the probe's mean free path is very large, so that its trajectory will not be blurred by interactions with the low-pressure environment.

\section{Conclusions}

With the goal of observing the gravitational potential of a delocalized quantum source, we have studied the possibility of fighting decoherence by freezing the state using the Zeno effect, and verifying the state using classical scattering of probe particles. While the Zeno freezing gives rise to a potential that could be achieved also by a classical mass distribution, we have argued that delocalization of quantum origin can be demonstrated in a controlled experiment. Finally, we estimated the feasibility parameters.

This estimate shows that our proposal is going to be as challenging as related ones, although some of the challenges to be met are different. For example, in proposals to observe gravity-mediated entanglement, localization due to decoherence places heavy constraints on the mass and delocalization scale of the source. In our case, decoherence bounds only the rate of the Zeno measurement, provided one is able to freeze a delocalized state (and we have only sketched how this could be done, leaving a detailed study for further study). Most likely, the experiment that will eventually demonstrate the delocalization of a gravity source has not yet been designed. At this early stage, it is important to explore ideas. The Zeno effect, either as used here or in a different form, might have a role in achieving the goal. 

\section*{Acknowledgments}

We thank \v{C}. Brukner, M. Carlesso, S. Pascazio, A. Pontin, O. Romero-Isart, H. Ulbricht, M. J. Romero, and R. Filip for discussions and comments.

We acknowledge financial support from the National Research Foundation and the Ministry of Education, Singapore, under the Research Centres of Excellence program.

 \bibliography{ZenoGravity}

\end{document}